# Resistive switching behaviors in vertically aligned $MoS_2$ films with Cu, Ag, and Au electrodes

*Shuei-De Huang, Touko Lehenkari, Topias Järvinen, Seyed Hossein Hosseini-Shokouh, Farzaneh Bouzari, Krisztian Kordas, Hannu-Pekka Komsa\**

**List of affiliations**

S.-D. Huang, T. Lehenkari, T. Järvinen, K. Kordas, H.-P. Komsa

Microelectronics Research Unit

Faculty of Information Technology and Electrical Engineering

University of Oulu

P. O. Box 4500, FI-90014 Oulu, Finland

E-mail: hannu-pekka.komsa@oulu.fi

S.-D. Huang, T. Lehenkari, T. Järvinen, K. Kordas, H.-P. Komsa

Infotech Oulu

University of Oulu

P. O. Box 4500, FI-90014 Oulu, Finland

S. H. Hosseini-Shokouh, F. Bouzari,

Department of Electronics and Nanoengineering

Aalto University

Tietotie 3 FI-02150, Finland.

**ORCID:**

Shuei-De Huang: https://orcid.org/0000-0001-9819-391X

Touko Lehenkari: https://orcid.org/0009-0007-2596-7030

Topias Järvinen: https://orcid.org/0000-0002-9256-748X

Seyed Hossein Hosseini-Shokouh: https://orcid.org/0000-0002-2356-7611

Farzaneh Bouzari: https://orcid.org/0009-0009-0512-7995

Krisztian Kordas: https://orcid.org/0000-0002-7331-1278

Hannu-Pekka Komsa: https://orcid.org/0000-0002-0970-0957

**Table of Contents**

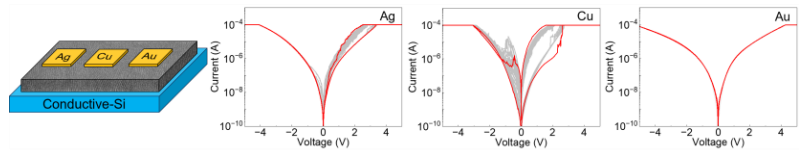

Vertically aligned MoS$_2$ memristors with Ag, Cu, and Au electrodes exhibit distinct switching behaviors on the same active material. Ag shows volatile switching, Cu enables stable non-volatile switching, and Au remains inert, demonstrating a simple approach for constructing diverse neuromorphic functions using a single platform.

**Abstract**

Neuromorphic computing circuits can be realized using memristors based on low-dimensional materials enabling enhanced metal diffusion for resistive switching. Here, we investigate memristive properties of vertically aligned $MoS_2$ (VA-$MoS_2$) films with three different metal electrodes: Ag, Cu, and Au. Despite having the same active material, all three metals show distinct switching behavior, which are crucial for neuromorphic computing applications: Ag enables volatile switching, Cu demonstrates stable non-volatile switching with retention over 2500 s, and Au shows no memristive response. Cu devices show abrupt resistance changes, and significant increase of copper content upon biasing, indicative of stable non-volatile switching based on filament formation and rupture. About 85% of Ag and Cu devices exhibit reliable memristor behavior. Our findings provide valuable insights into the memristive switching mechanism in VA-$MoS_2$ and present a promising avenue for facile fabrication of neuromorphic circuits by employing a set of different metals on a single active material.

## 1. Introduction

Research on artificial intelligence and machine learning has received widespread attention in recent years, but the implementations on the existing von Neumann computing architecture are highly inefficient[1–4], because data needs to be continuously transferred between the processor and memory, causing high energy consumption and latency. This has motivated research on novel processor architectures, such as in-memory computing and neuromorphic computing. In in-memory computing, the memory and processor are co-located on the chip, while still relying on the traditional CMOS logic[5,6]. Neuromorphic computing presents a more drastic departure, which aims at more closely emulating the way information is processed in brains[7,8]. At the core of this approach lie the fundamental building blocks of the nervous system: neurons and synapses. Neurons are computational units that process and transmit signals[9], while synapses are the connections between neurons and modulate signals' transmission strength. The strength of these synaptic connections can change over time, which is crucial for learning and memory formation[10,11]. Moreover, while the information is transmitted in the form of electrical signals, the information is encoded in the form of spikes instead of continuous analog signals.

While the neuromorphic functionalities can be emulated in CMOS logic[12–16], more efficient implementations (in terms of energy consumption and die space) are required. Memristors are considered to have the potential to function as synaptic elements in neuromorphic computing systems due to their ability to emulate biological synapses and neurons[17–20]. Memristors (from "memory resistors") are circuit elements whose resistance depends on the history of currents passed through the device[21–26]. Recent investigations have highlighted the promising capabilities of memristors in these applications, particularly their role in mimicking the essential features of synaptic plasticity depending on the volatility of the resistance changes. For instance, volatile memristors have been effectively utilized as artificial neurons[3,27], displaying spiking behavior like the biological neurons, while non-volatile memristors have been demonstrated as effective synaptic devices[28,29], capable of storing and modulating synaptic weights over time. These findings suggest that memristors could significantly enhance the performance and scalability of neuromorphic systems[30,31].

First memristor devices were fabricated using metal oxides, such as $TiO_2$[32], but recently 2D materials, especially $MoS_2$, have garnered significant attention in the development of advanced memristors due to their unique properties[33–37]. Compared to traditional metal oxide memristors, 2D materials like $MoS_2$ can significantly enhance the operating speed and energy efficiency of these devices. While monolayer $MoS_2$ memristors achieve ultrafast switching due to the ultrathin active region, they also suffer from excessive leakage currents as a single point defect can create a conductive path across the entire layer. On the other hand, lateral multi-layered $MoS_2$ memristors provide improved reliability but the relatively large electrode spacing dictated by lithographic resolution limits the switching speeds. These challenges have driven interest toward vertically aligned $MoS_2$ (VA-$MoS_2$), where the layered structure enables rapid metal migration through interlayer spaces while maintaining a compact and scalable architecture. While the memristive behavior is generally ascribed to metal ion migration, either by filling the vacancies or by forming conductive filaments[38–40], the reports regarding the behavior of different metals are scattered. First, memristors fabricated with vertically grown $MoS_2$ and Ag electrode have demonstrated contrasting behaviors in different studies. In study by Dev et al.[33], the devices exhibited abrupt volatile memristive behavior, where the resistance states are temporary, returning to high-resistance states immediately after voltage is removed. Conversely, Ranganathan et al.[34] reported non-volatile behavior in similar devices, although with noticeably larger switching voltage. Second, Au was successfully used in the case of monolayer memristors (or "atomristors"), whereas Au is commonly used as a passive electrode in $MoS_2$ electronics. Third, Cu[41–44] have been shown to readily intercalate into $MoS_2$, suggesting potential for memristive behavior. However, to date, no studies have been reported investigating the memristive properties of VA-$MoS_2$ with Au or Cu top electrodes. This discrepancy in observed memristive behaviors highlights the need for a comparative study among different metals to understand the factors influencing the volatile and non-volatile characteristics of $MoS_2$-based memristors.

In this paper, we fabricate VA-$MoS_2$-based memristors with three different active metals, Ag, Au, and Cu. We find that all three metals show different behavior: Ag yields volatile memristors, Cu non-volatile memristors, and no memristive effect is observed with Au. We perform careful material characterization on all three devices and correlate them with the electrical characteristics to understand how different metal electrodes influence the switching mechanisms and stability of memristors.

2. Results and discussion

To synthesize the VA-$MoS_2$ films, 20nm of Mo was deposited on heavily p-doped Si substrate and sulfurized at 700°C. To directly assess the structure of thermally sulfurized Mo on Si substrates, a cross-sectional lamella was cut and polished by FIB for subsequent TEM imaging. As shown in Figure 1a, the on-Si grown thin film with a thickness of ~50 nm has a well-oriented layered crystal structure, with a perpendicular alignment in reference to the surface of the substrate. The spacing of the layers is measured to be 6.2 Å by averaging the distance between multiple periods in different locations of the lattice. This value is in good agreement with bulk 2H-$MoS_2$ (6.15 Å)[45]. In-plane XRD (Figure 1d) shows only the (002) reflection at $2\theta = 14.08°$, whereas the diffraction pattern acquired by grazing incidence analysis lacks that, clearly proving the vertical alignment of $MoS_2$ on Si[46,47]. These results on the structure of thin films of sulfurized Mo are similar to those reported earlier by several groups[33,34,46,48] indicating the synthesis method is robust and well repeatable.

Raman spectrum in Figure 1c shows the characteristic LA(M), $E_{1g}$, $E_{2g}$ and $A_{1g}$ vibration modes of $MoS_2$ at 226.1 $cm^{-1}$, 285.6 $cm^{-1}$, 384 $cm^{-1}$ and 410.4 $cm^{-1}$, respectively. LA(M) peak, arising from LA phonons at the M point in the Brillouin zone, is activated by defects[49–52], and thus very low LA(M)/$A_{1g}$ intensity ratio (~0.06) indicates the good quality of the $MoS_2$ film[50]. The large separation (26.4 $cm^{-1}$) of in-plane ($E_{2g}$) and out-of-plane ($A_{1g}$) vibration mode peaks corresponds to bulk-like layered structure[53], whereas the small ratio of their intensities (~0.25) indicates that mainly the edges of $MoS_2$ are exposed on the surface, which is typical of vertically aligned $MoS_2$[48,54]. Finally, AFM topography image in Figure 1b shows that the $MoS_2$ surface is smooth and even, with an average roughness of Rq = 1.49 nm.

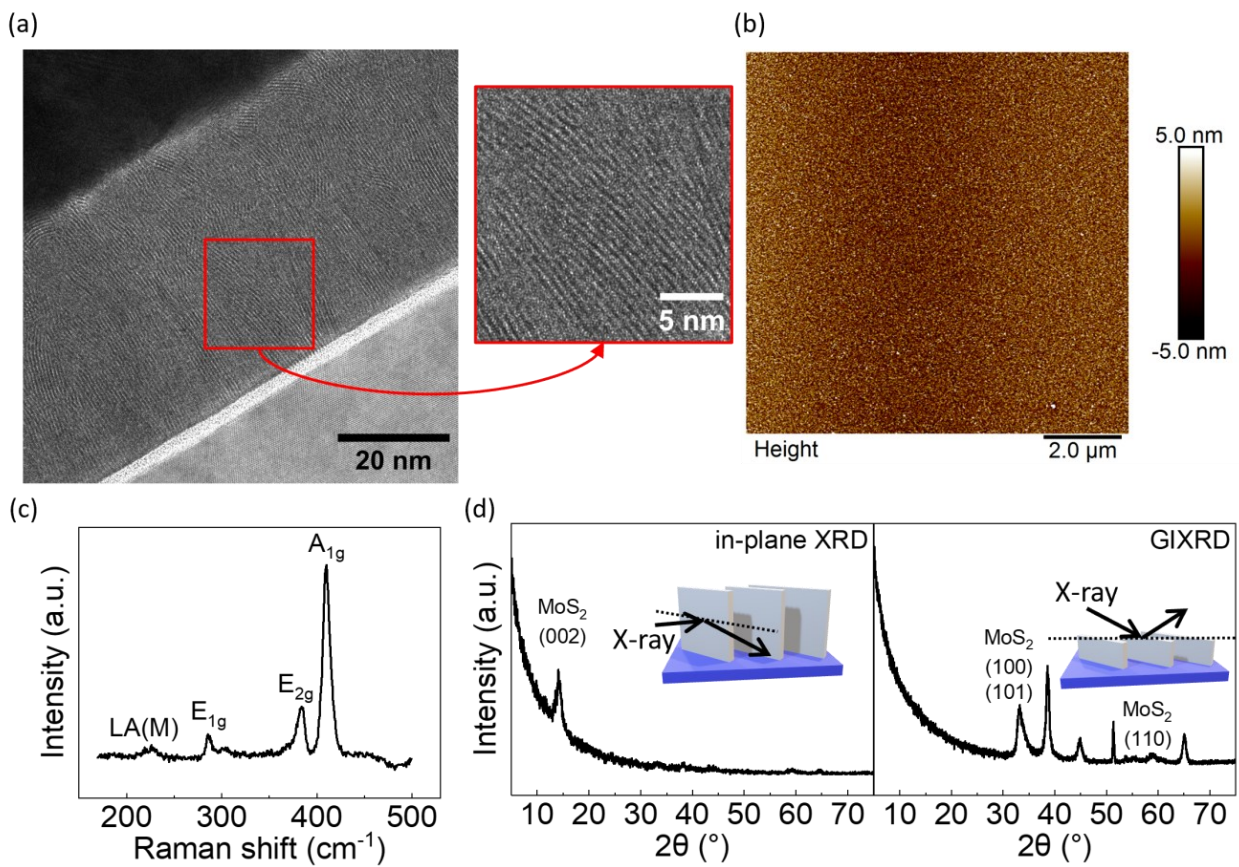

Figure 1. Characterization of VA-$MoS_2$ films. (a) TEM cross-section image and a zoom-in to the $MoS_2$ layer to highlight the vertically aligned layers. (b) AFM topography image from VA-$MoS_2$. (c) Raman spectra of the film with the peak assignments. (d) In-plane XRD and GIXRD pattern with selected peak assignments. The measurement geometry is illustrated in the insets.

Next, three types of devices were fabricated by depositing different metal thin film top electrodes, including Ag, Cu, and Au. Their I-V characteristics were measured as illustrated in Figure 2a. Figure 2b shows multiple consecutive current-voltage sweeps of Ag/VA-MoS$_2$/Si device measured between –5 V and +5 V with step size 10 mV, step delay 10 ms, and a compliance current of 100 µA. It can be observed that upon positive scans starting at 0 V, the device gradually switches from a high resistance state (HRS) to a low resistance state (LRS). On a reverse scan the current starts to decay at ~2 V and shows clear hysteresis. Scans from 0 V to –5 V and back are nearly identical and resemble the original HRS, i.e., the device has turned off already before applying negative voltage and thus exhibits volatile switching. Devices are stable for 30 cycles, and the on-off ratio is about 10. About 90% of the not-short-circuited Ag/VA-MoS$_2$/Si devices show similar memristor behavior (I-V characteristics from 23 devices are collected in Figure S2 in SI). The volatile switching behavior is consistent with the devices reported by Dev et al[33], although at a higher voltage and less abrupt switching, which could be due to a thicker MoS$_2$ film and different bottom electrode. Furthermore, our devices do not require an electroforming step, unlike the devices reported by Ranganathan et al[34].

In contrast to Ag/VA-MoS$_2$/Si, devices with copper top electrodes (Cu/VA-MoS$_2$/Si) show dramatically different current-voltage characteristics (Figure 2c). During the reverse voltage sweep, the device stays in LRS, until a voltage of –1 V is applied, at which a reset to the HRS takes place, i.e. non-volatile switching is observed. Devices are stable for 30 cycles, and the on-off ratio is about 50. About 80% of the not-short-circuited Cu/VA-MoS$_2$/Si devices show similar memristor behavior (I-V characteristics from 33 devices, with –5 or –10 V reset voltage, are collected in Figure S3). Figures S2-S3 also indicate the age of the sample during characterization. The samples do not exhibit qualitative changes to the switching behavior when measured after a year of storage. Interestingly, the resistive switching characteristics tend to exhibit small improvement over time, especially in the case of non-volatile Cu devices. The possible reason for the improvement is the intercalation of water and air inside the active region from the sides of the sample, whose reduction at the inert electrode help compensate for the metal oxidation at the active electrode, as proposed in the case of h-BN and metal oxide memristors[55,56].

For reference, we also tested devices using Au as the top electrode. As can be seen in Figure 2d, multiple subsequent I-V measurements result in nearly identical plots, of which each is exempt from any hysteresis, i.e., the devices do not exhibit any switching behavior.

We also investigated the switching behavior of Ag/VA-MoS$_2$/Si and Cu/VA-MoS$_2$/Si devices under a compliance current of 500 µA. This resulted in a distinctly different behavior in the case of Ag/VA-MoS$_2$/Si devices, with approximately 50% of the switching cycles exhibiting non-volatile behavior (highlighted as green curves in Figure 2e), while the remaining cycles still retained volatile switching characteristics. In contrast, the Cu/VA-MoS$_2$/Si devices consistently demonstrated non-volatile switching behavior even under the higher compliance current. However, to avoid a permanent transition to the LRS, the voltage range must be restricted to −3 V to 3 V, as shown in Figure 2f. These observations indicate that the compliance current is a critical factor influencing the switching behavior. Nevertheless, the properties of the top electrode metal also play a key role. Notably, Cu tends to form stronger and more stable conductive filaments compared to Ag.

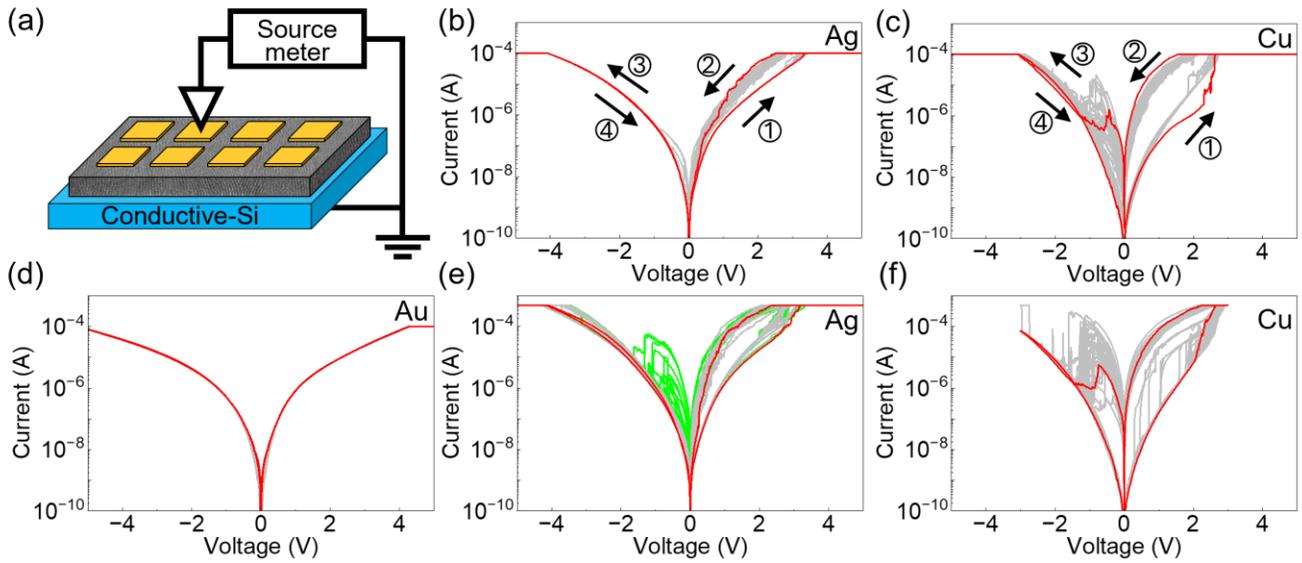

Figure 2. I-V characteristics of metal/VA-MoS$_2$/Si devices. (a) Illustration of the measurement setup. The indicated voltage is applied to the top metal electrode. (b-d) 30 consecutive I-V sweeps with compliance current of 100 μA on the (b) Ag/VA-MoS$_2$/Si, (c) Cu/VA-MoS$_2$/Si, and (d) Au/VA-MoS$_2$/Si devices. (e-f) 30 consecutive I-V sweeps with compliance current of 500 μA on the (e) Ag/VA-MoS$_2$/Si, (f) Cu/VA-MoS$_2$/Si. The red curves highlight the first sweep, and the non-volatile cycles in (e) are highlighted with green color, while the rest are colored gray. The numbered arrows indicate the sweep sequence.

To understand the underlying reasons for the different memristive characteristics of Ag/VA-MoS$_2$/Si and Cu/VA-MoS$_2$/Si devices, energy dispersive X-ray spectroscopy (EDS) mapping of Ag and Cu was carried out on FIB cross-sectioned devices before any voltage sweep (i.e. pristine samples) and after setting them to the low-resistance states. For the latter, we applied +5 V with a compliance current of 100 μA for 5 min to fully set both Ag/VA-MoS$_2$/Si and Cu/VA-MoS$_2$/Si devices and then opened the circuits. Under such conditions, according to the I-V measurements shown in Figure 2, the Ag/VA-MoS$_2$/Si is supposed to return to its high-resistance state, whereas the Cu/VA-MoS$_2$/Si is expected to remain in its low-resistance state. The corresponding Ag elemental maps (Figure 3b) and integrated concentration profiles (Figure 3c) of two identical devices show that the concentration profile of Ag in the $MoS_2$ film is practically the same for the pristine sample and for the biased sample. The result implies that, first, the applied voltage cannot induce extraction of a significant concentration of Ag atoms from the electrode to the film and therefore the memristive behavior likely involves Ag atoms already trapped within the film. Second, any movement of $Ag^+$ ions (and possibly filament formation) induced by the field is negated after the field is removed, likely by thermal diffusion and/or Joule heating of the filament. These results are fully consistent with the volatile behavior observed in the I-V measurements. However, the behavior of the Cu devices is markedly different. The concentration of Cu in $MoS_2$ after biasing is significantly higher than in an identical pristine chip without biasing. This observation gives direct evidence that Cu is extracted from the top electrode and inserted into the $MoS_2$ film, and that the Cu stays in the film after the voltage is removed and thus exhibiting non-volatile behavior in I-V measurements. However, as we cannot directly observe Cu filaments, it gives only indirect information about the physical mechanism behind the resistive switching. Finally, it is worth noting that the EDS maps show negligible Ag/Cu diffusion into the $Si/SiO_2$ layers and no $Si/MoS_2$ interdiffusion, further evidencing that the memristive switching arises from metal migration within the VA-$MoS_2$ film.

Recent first-principles calculations on interstitial metal defect formation and migration in $MoS_2$ by Lehenkari et al. support these findings[57]. Both Ag and Cu were found to predominantly exist in the +1 charge state, enabling their manipulation with an electric field. Formation of intercalated $Ag^+$ ions extracted from the metal electrodes is expected only in very small concentrations, mainly within grain boundaries, where their rapid diffusion can facilitate swift filament formation and breaking. These likely account for the absence of additional Ag concentration in the EDS data and the volatile nature of memristive switching observed in devices. In contrast, $Cu^+$ ions were found to form in much larger concentrations, and primarily within the interlayer spaces of the bulk material. In this environment, $Cu^+$ ions diffuse more slowly but exhibit greater stability. This explains the higher Cu concentrations observed in EDS data and the nonvolatile switching behavior, attributed to the formation of more stable filaments. Lastly, Au was also examined in this study, but high formation energies (>1.5 eV) make the presence of gold in significant concentrations unlikely. This is consistent with our observations, as memristors utilizing Au top electrodes did not exhibit memristive switching.

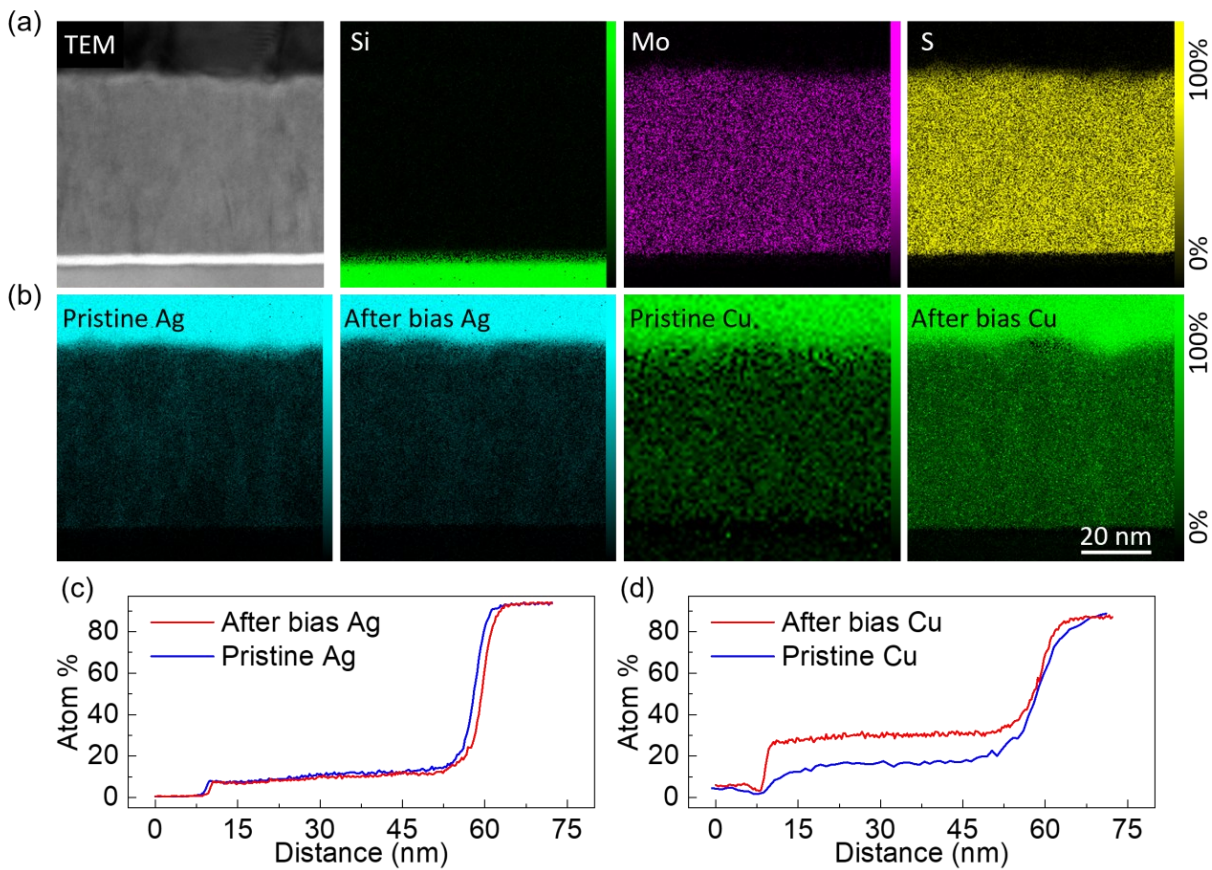

Figure 3. EDS maps from the samples before and after biasing. (a) TEM image and EDS maps (in atom %) for Si, Mo, and S from pristine Ag sample. (b) Active metal distribution before and after applied bias. (c,d) Integrated concentration profiles from the Ag and Cu samples. The distance is measured from the bottom of the image.

We conducted data retention measurements for both Ag/VA-MoS$_2$/Si and Cu/VA-MoS$_2$/Si devices. First, a +5 V bias was applied for 30 seconds to set the devices to the LRS. After the bias was removed, a 0.5 V bias was applied every second to verify the device state. For the Ag/VA-MoS$_2$/Si device (Figure 4a), the device immediately switched to the HRS after the bias was removed. To confirm the complete return to HRS state, we reset the device using a –5 V bias and then resumed the same measurement procedure, applying a 0.5 V bias every second. The results consistently showed that the device remained in the HRS (Figure S4). In contrast, the Cu/VA-MoS$_2$/Si device (Figure 4b) stays in the LRS for more than 2500 seconds; such non-volatile behavior is consistent with our I-V measurements. We repeated the data retention measurements multiple times, varying the set time to maximize retention duration (Figure 4c). Interestingly, we found that the retention time gradually increased with repeated voltage applications, regardless of the initial set time.

In addition to the retention time, we were also interested in the dynamics of the initial switch to LRS. To investigate this, we connected our memristor devices in series with a resistor, using a sourcemeter to apply a bias and an oscilloscope to measure the voltage across the load resistor ($R_L$ = 22 kΩ). When the voltage across the resistor increases, it indicates that the memristor device has switched to the LRS. Upon applying the bias, the voltage across the resistor ($V_R$) for the Ag device (Figure 4f) gradually increased and stabilized at approximately 3 V, corresponding to about 2 V across the memristor device. The absence of abrupt voltage changes, in combination with the very fast volatile behavior described above, suggests that the switching may not arise from filament formation, but has another origin, e.g. modification of the Schottky barriers upon Ag migration[58]. On the other hand, Cu devices (Figure 4g) show abrupt resistance changes that can be ascribed to formation and rupture of copper filaments. We note that in our measurement setup the voltage across the memristor is automatically regulated to reside close to the switching voltage. As the device switches to LRS, the voltage across it drops according to the voltage divider rule, preventing the formation of thicker filament (or additional filaments) and ultimately enabling filament rupture/dissolution. Consequently, from these measurements, we can extract a more precise value of 2.4 V for the voltage that is required for switching on the device.

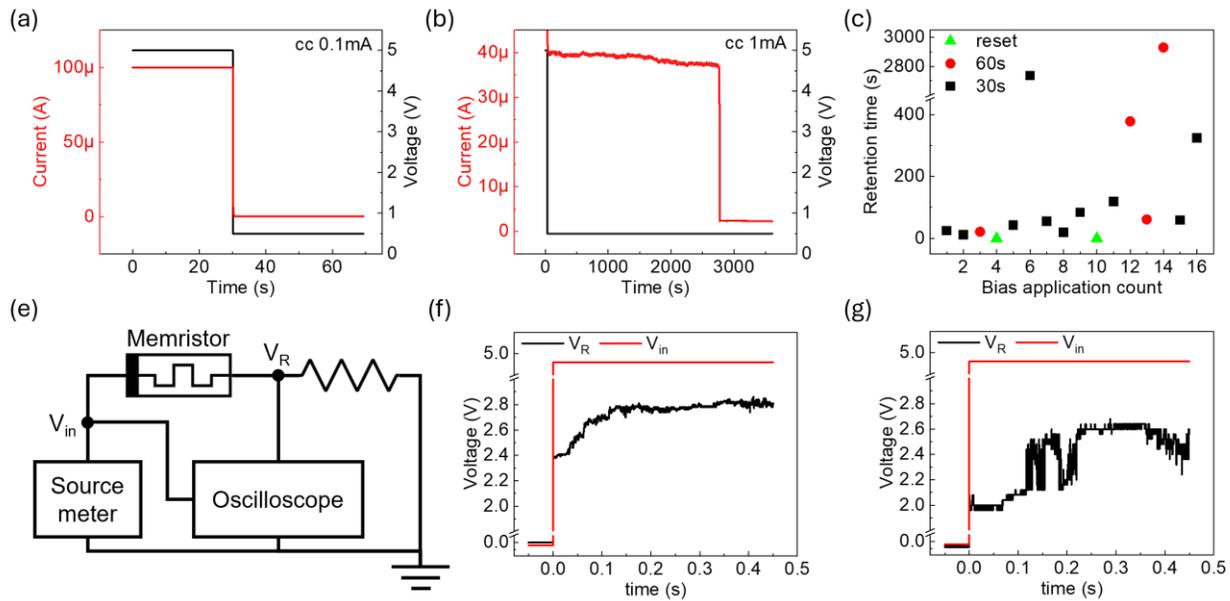

Figure 4. Memristive switching dynamics. (a,b) Measured current and applied voltage as a function of time of (a) Ag and (b) Cu devices, from which the retention time of memristor devices can be extracted. (c) Measured retention times from a single device upon repeated application of set bias. Two different set durations were used as indicated by the marker color. (e) Sketch of the measurement setup for probing the switch to low-resistance state. (f,g) Evolution of the voltage over the load resistor $V_R$ as a function of time from (f) Ag and (g) Cu devices.

## 3. Conclusions

We systematically investigated the memristive behavior in vertically aligned VA-MoS$_2$ films with three different top metal electrodes: Ag, Au, and Cu. Our results reveal distinct differences in the performance and functionality of the memristors depending on the choice of the metal electrode. The Ag/VA-MoS$_2$ device exhibited volatile switching behavior, consistent with previously reported studies. EDS indicated no persistent changes in the Ag distribution before and after biasing. In contrast, Cu/VA-MoS$_2$ device demonstrated non-volatile switching with significant increase in the Cu concentration inside the VA-MoS$_2$ film upon biasing. In addition, the resistance changes are abrupt, suggesting filament-type switching mechanism. To the best of our knowledge, this is the first report of Cu/VA-MoS$_2$ memristor. Finally, the Au/VA-MoS$_2$ device did not exhibit any memristive switching. This study provides insights into the mechanisms underlying the switching behavior of VA-MoS$_2$-based memristors and paves the way for further research into optimizing metal-semiconductor combinations for neuromorphic computing applications. Importantly, our results demonstrate that a wide variety of neuromorphic computing functionalities can be obtained with a single active material by simply depositing different metals as the top electrode, thus offering a promising platform for facile fabrication of integrated neuromorphic circuits.

## 4. Experimental section

### 4.1. Synthesis of VA-MoS$_2$ and device fabrication

MoS$_2$ film growth was carried out on heavily doped silicon (p-type boron doping <100> R<0.005 ohm-cm) substrates, which were then used as bottom electrodes of our devices. For MoS$_2$ film synthesis, first, a layer of 20 nm Mo was deposited on the cleaned (acetone and IPA ultrasonic bath) Si surface by DC sputtering (Torr International PVD system, see Table S1), followed by a thermal sulfurization process. AFM image in Figure S1 shows that the Mo seed layer has small roughness of Rq = 0.89 nm. The sulfurization was achieved by placing a sulfur powder loaded (1 g) alumina boat and the Mo coated Si substrates loaded alumina boat into a horizontal tube furnace (Thermo Scientific Thermolyne with 2-inch quartz tube) so that the chips were in a downstream position (right after the sulfur powder loaded boat). After evacuating the reactor to a base pressure below 1 Torr with an oil pump, the tube was flushed with $N_2$. This was repeated three times to minimize $O_2$ and water vapor content in the reactor. Next, the tube furnace was heated to 700°C with a rate of 109°C/min and the temperature was kept for one hour, while passing through 400 sccm $N_2$ to carry the evaporated sulfur over the chips. Finally, while maintaining the $N_2$ flow, the reactor was left to cool down naturally to room temperature before removing the sulfurized samples.

The top electrodes (each with a size of 500 × 500 µm$^2$) were deposited by sputtering (Torr International PVD system, see Table S1) metal stacks of Cu/Au (100/100 nm), Ag/Au (100/100 nm), and Au (100 nm) onto the $MoS_2$ surface through a laser-cut alumina shadow mask (thickness of 250 µm).

**4.2. Characterization**

Devices cross-section lamellae for TEM analysis (imaging and energy dispersive X-ray spectroscopy elemental mapping, JEOL JEM-2200FS EFTEM/STEM) were prepared using a focused ion beam (FEI Helios DualBeam FIB-FESEM/STEM). Grazing incidence X-ray diffraction (GIXRD) (Rigaku SmartLab 9 kW, Cu Kα-radiation, 45kV, 200mA, Parallel Beam, incidence angle ω 2°, scan speed 2.0169°/min, step width 0.02°, scan range 2θ from 5° to 90°) and in-plane X-ray diffraction (in-plane XRD) (incidence angle ω 0.2°, Goniometer axis positions 2θ 5°, scan speed 3.0251°/min, step width 0.04°, scan range 2θχ from 5° to 90°) were employed to verify the growth orientation of $MoS_2$.

Raman spectroscopy measurements were performed using a WITec alpha300 RA+ system, equipped with a 532 nm laser (at 1 mW), a Newton Andor EMCCD detector and a 100x Nikon CF Plan objective (NA = 0.95).

Scanning electron microscopy (SEM, Zeiss Sigma FESEM) and atomic force microscopy (AFM, Bruker Multimode 8) were employed to evaluate the surface topology and roughness of the VA-$MoS_2$. The AFM measurements were conducted using PFTUNA probes (Bruker) in PeakForce QNM mode. I-V characterization was carried out using a LabView controlled KEITHLEY 2636A source meter. The results shown in Figure 2 are from the aged samples, which resulted in slightly improved characteristics, as discussed in the main text, while all the material characterization was done on fresh samples.

**CRediT authorship contribution statement**

**Shuei-De Huang:** Design, Methodology, Validation, Formal analysis, Investigation, Writing - Original Draft, Writing - Review & Editing. **Touko Lehenkari**: Validation, Formal analysis. **Seyed Hossein Hosseini-Shokouh**, and **Farzaneh Bouzari**: Validation, Formal analysis, Review and Editing. **Krisztian Kordas**: Supervision, Formal analysis, Funding acquisition, Resources, Writing - Review & Editing. **Hannu-Pekka Komsa**: Supervision, Formal analysis, Funding acquisition, Resources, Writing - Review & Editing.

**Declaration of Competing Interest**

The authors declare that they have no known competing financial interests or personal relationships that could have appeared to influence the work reported in this paper.

**Acknowledgements**

We acknowledge funding received from Infotech Oulu under project "Memristors and neuromorphic sensors from vertically aligned layered materials", Kvantum Institute under project "Robust 2D materials for sensors, photo and electrocatalysis" and FARIA through project "Ultrafast detectors and rectifiers based on vertically aligned 2-dimensional materials" at the University of Oulu. S.H. Hosseini-Shokouh acknowledges funding from Research Council of Finland (RCF) through project Order2Chaos/Grant No. 13357170. We also would like to thank Marcin Selent, Esa Heinonen and Sami Saukko (Centre for Material Analysis, University of Oulu) for their invaluable help in XRD characterization, FIB processing and TEM analysis of samples.

**Appendix A. Supplementary material**

Supplementary material

**Data Availability**

Data will be made available on request.

# Supporting Information

# Resistive switching behaviors in vertically aligned MoS$_2$ films with Cu, Ag, and Au electrodes

*Shuei-De Huang [a,b], Touko Lehenkari [a,b], Topias Järvinen [a,b], Seyed Hossein Hosseini-Shokouh [c],*

*Farzaneh Bouzari [c], Krisztian Kordas [a,b], Hannu-Pekka Komsa [a,b,]\**

[a] Microelectronics Research Unit, Faculty of Information Technology and Electrical Engineering, University of Oulu, P. O. Box 4500, FI-90014 Oulu, Finland

[b] Infotech Oulu, University of Oulu, P. O. Box 4500, FI-90014 Oulu, Finland

[c] Department of Electronics and Nanoengineering, Aalto University, Tietotie 3 FI-02150, Finland.

\***Corresponding Authors**

E-mail: hannu-pekka.komsa@oulu.fi

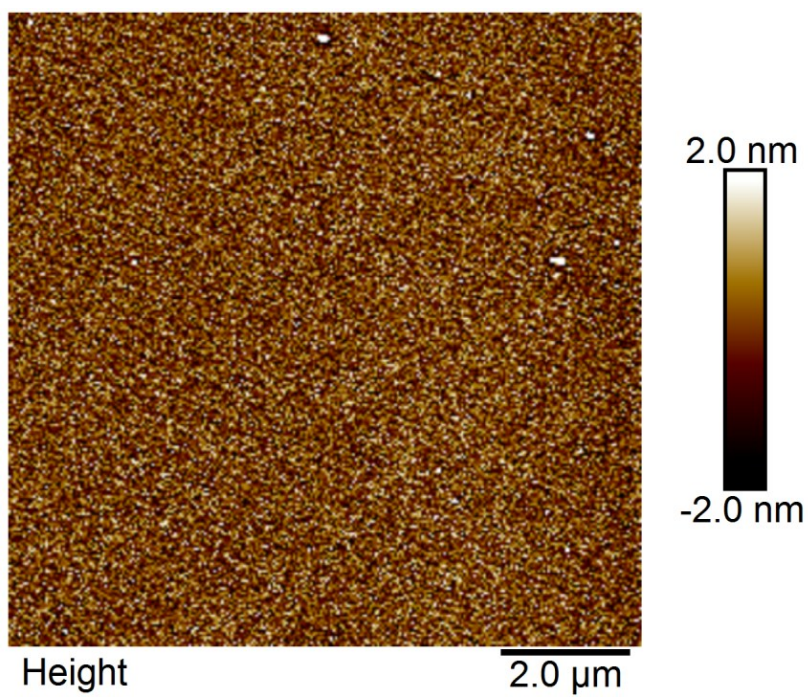

Figure S1. AFM topography image of Mo seed layer prior to sulfurization, with Rq = 0.886nm.

|  | Mo | Ag | Cu | Au |
|---|---|---|---|---|
| Voltage | 340 V | 440 V | 447 V | 421 V |
| Current | 130 mA | 160 mA | 213 mA | 125 mA |
| Deposition rate | 0.5 Å/s | 2.2 Å/s | 1.6 Å/s | 1.8 Å/s |

Table S1. Details of metal deposition.

**Additional electrical characterization of memristor devices**

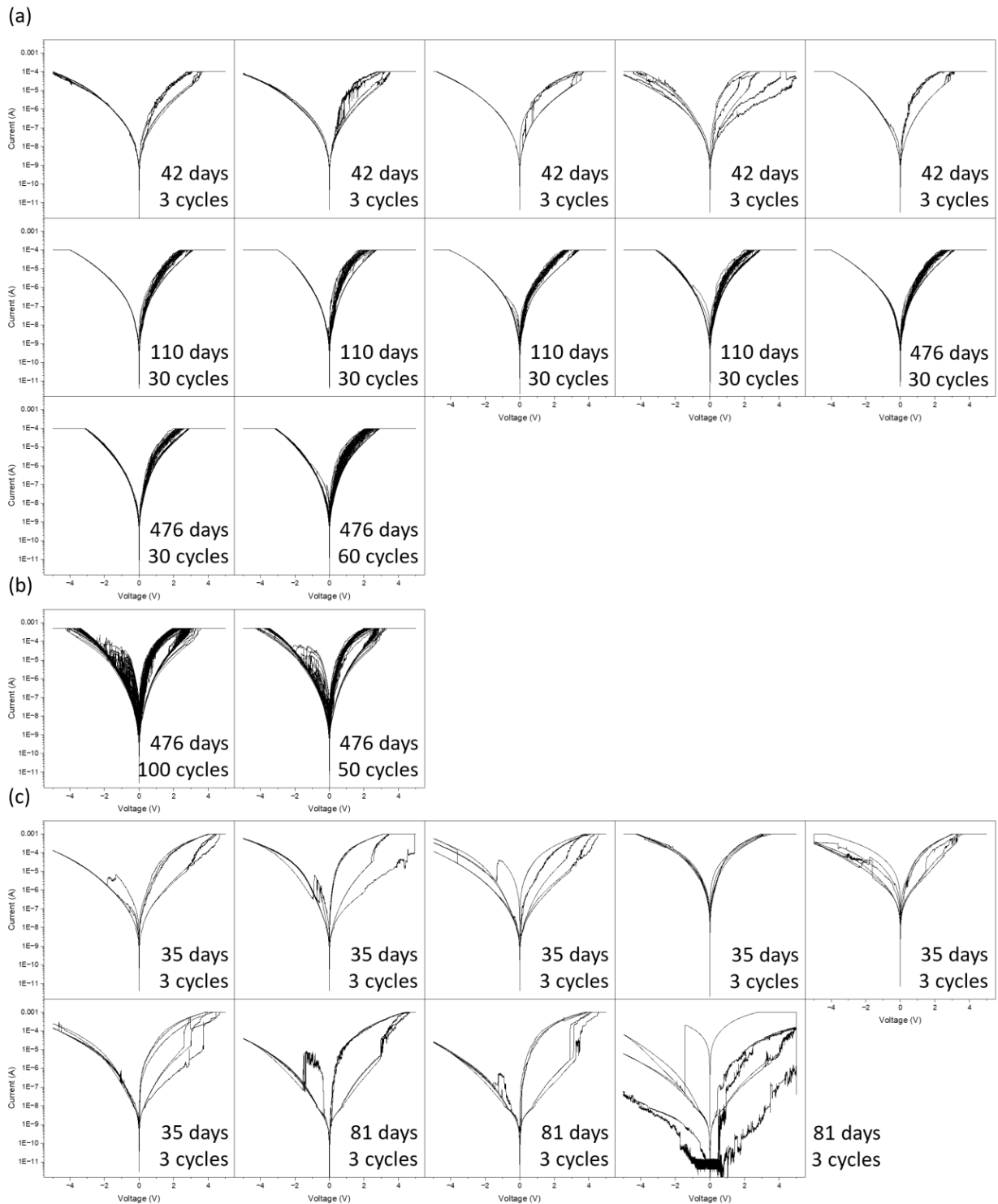

Figure S2. I-V characteristics of a set of Ag/VA-MoS$_2$/Si memristor devices under different compliance currents: (a) 100 µA, (b) 500 µA, and (c) 1 mA.

(a)
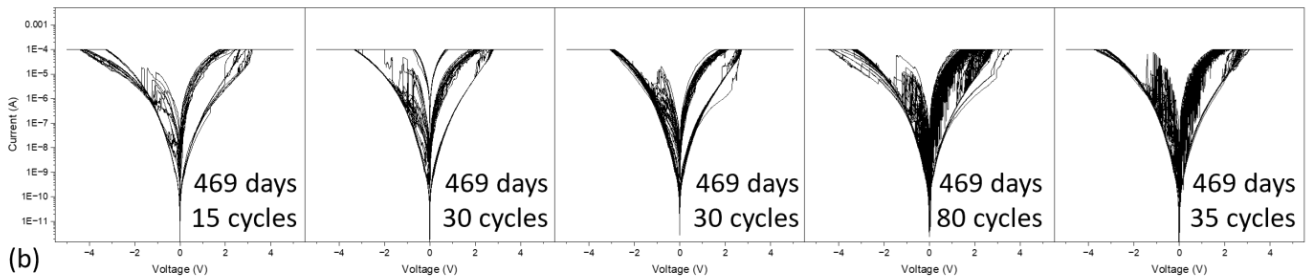
(b)
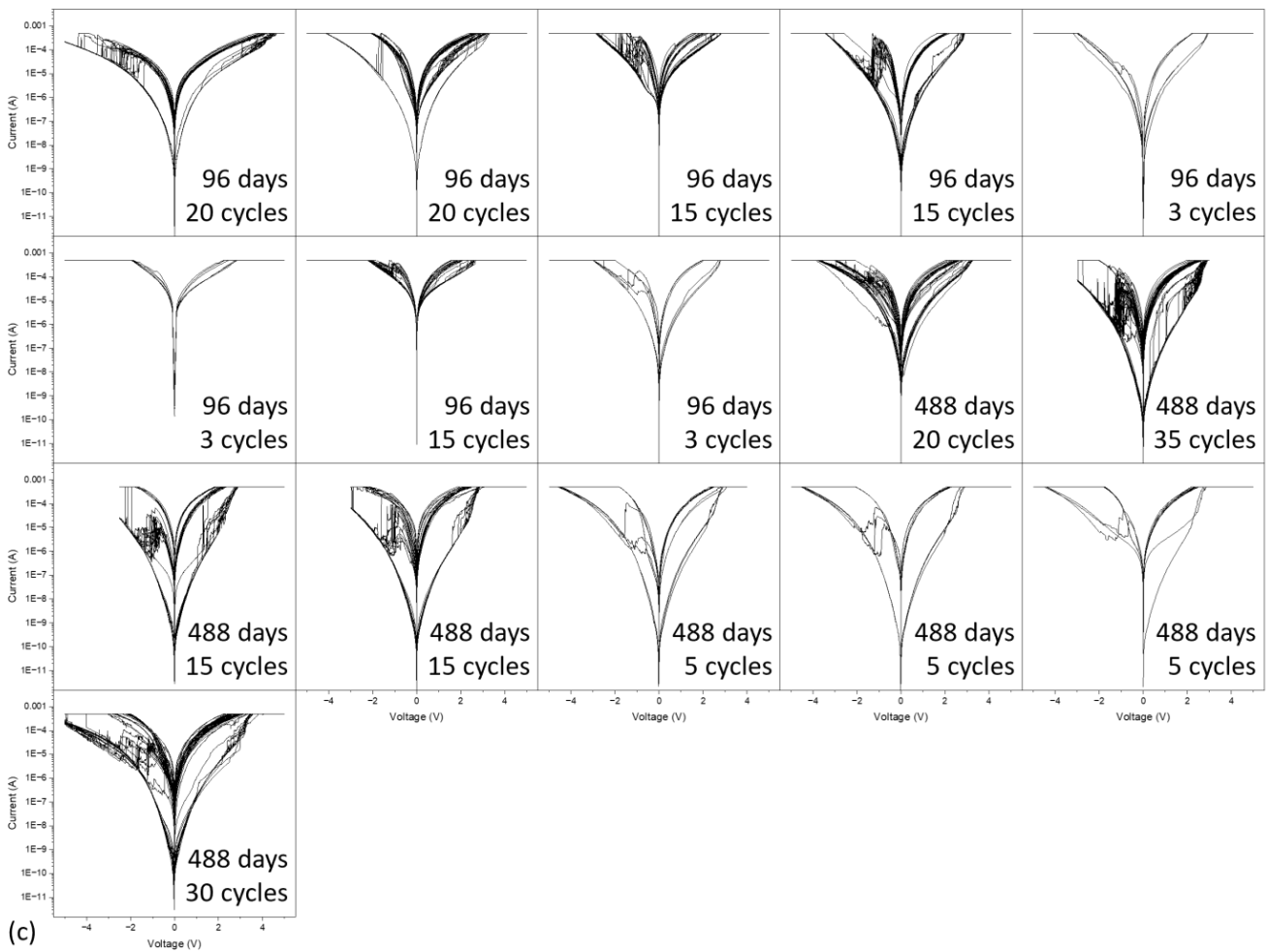
(c)
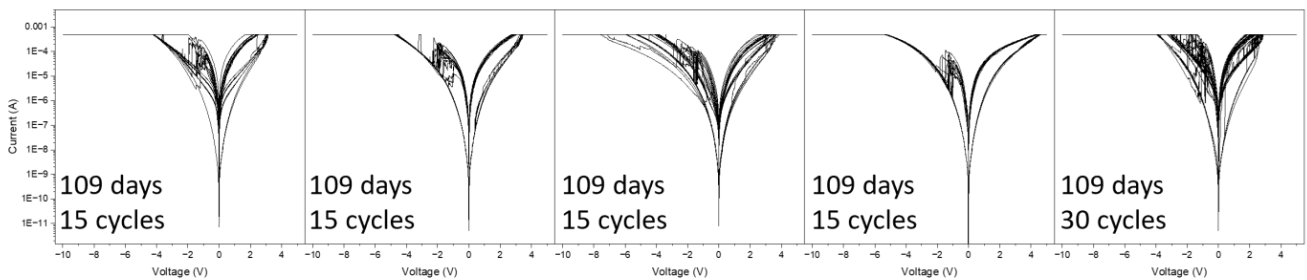

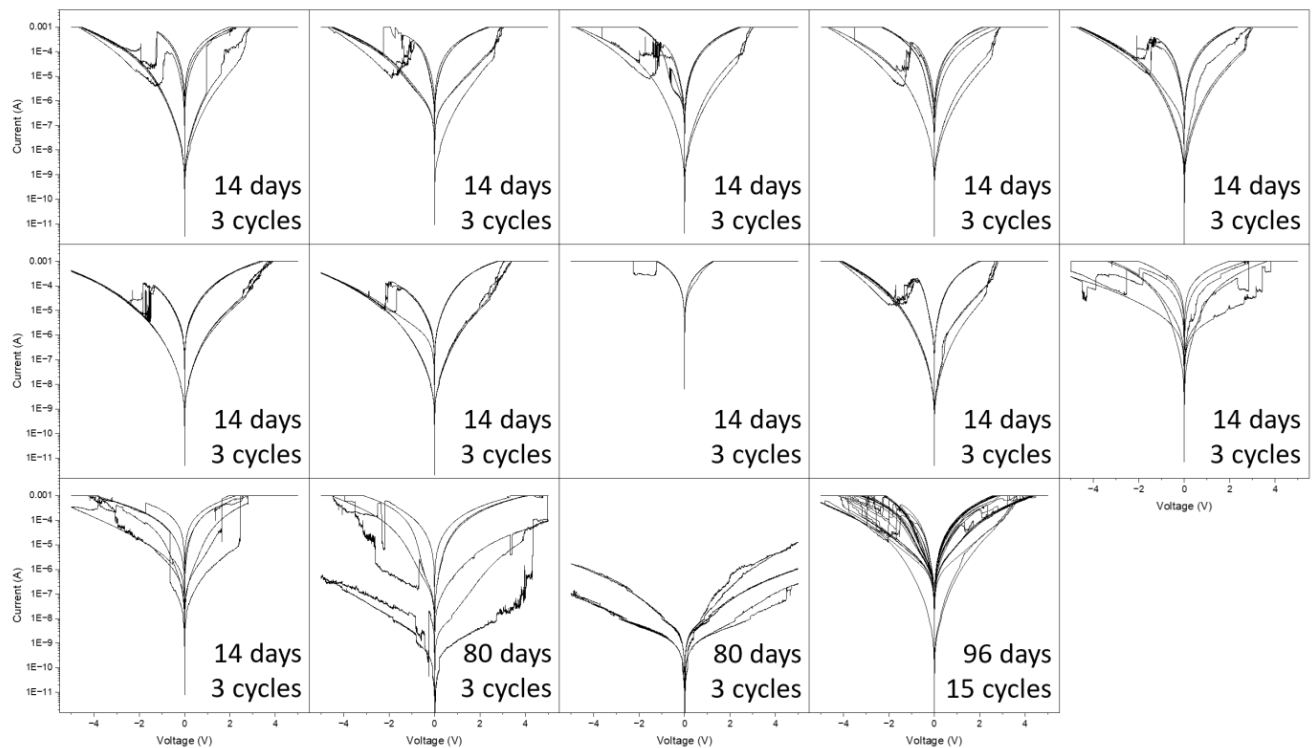

Figure S3. I-V characteristics of a set of Cu/VA-MoS$_2$/Si memristor devices, under different compliance currents: (a) 100 µA, (b-c) 500 µA, and (d) 1 mA.

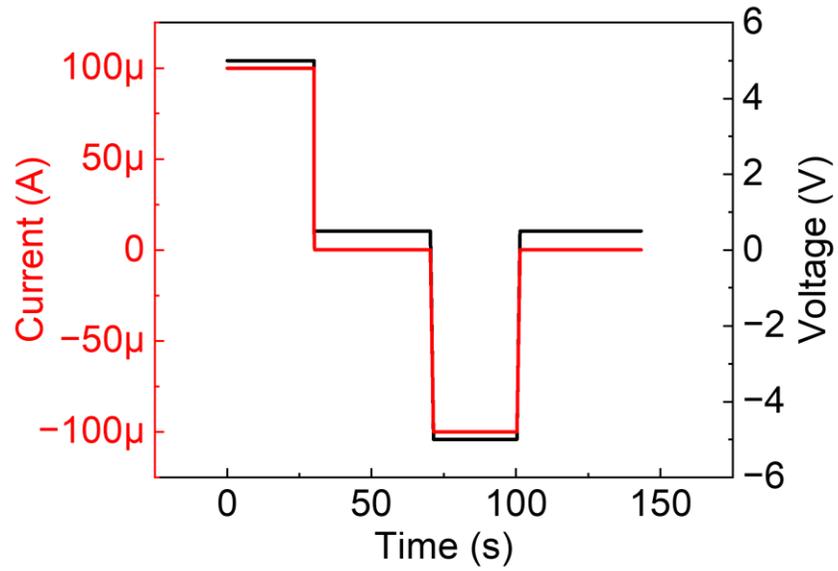

Figure S4. Data retention measurement for Ag/VA-MoS$_2$/Si device, as in Figure 4a, but also including a reset pulse, thus demonstrating that the high resistance state before and after the reset pulse is the same.